\documentclass[preprint,12pt,times]{elsarticle}
\usepackage{amsmath,amssymb}
\usepackage{booktabs}
\usepackage{tabularx,array}
\usepackage{graphicx}
\usepackage{multirow}
\usepackage{siunitx}
\usepackage{microtype}
\usepackage{xurl}
\usepackage{placeins}
\usepackage{float}
\usepackage{hyperref}

\hypersetup{colorlinks=true,allcolors=blue!55!black,
pdftitle={InletScope: Multiscale pressure-based inlet unstart monitoring under changing sensor configurations},
pdfauthor={Jiaming Liu, Fenlou Zhai, Ao Xue, Liangliang Tang, Cheng Ding, Daoqiang Zhang}}
\journal{Acta Astronautica}
\begin{document}
\begin{frontmatter}

\title{InletScope: Multiscale pressure-based inlet unstart monitoring under changing sensor configurations}
\author[ai]{Jiaming~Liu}
\author[aeac]{Fenlou~Zhai}
\author[aeac]{Ao~Xue}
\author[cepe]{Liangliang~Tang}
\author[ai]{Cheng~Ding}
\author[ai]{Daoqiang~Zhang\corref{cor1}}
\ead{dqzhang@nuaa.edu.cn}
\cortext[cor1]{Corresponding author.}
\affiliation[ai]{organization={College of Artificial Intelligence, Nanjing University of Aeronautics and Astronautics},
city={Nanjing},postcode={211106},country={China}}
\affiliation[aeac]{organization={Aero Engine Academy of China, Aero Engine Corporation of China},
city={Beijing},postcode={101304},country={China}}
\affiliation[cepe]{organization={College of Energy and Power Engineering, Nanjing University of Aeronautics and Astronautics},
city={Nanjing},country={China}}

\begin{abstract}
Supersonic inlets capture and compress air for aeroengines, and unstart can disrupt the air supply and compromise propulsion-system stability. Pressure-based monitoring must accommodate flow responses that vary with operating condition and measurement location. InletScope combines shared multiscale pressure encoding, learned scale fusion and sensor-probability averaging, allowing measurement sets to change without retraining. Separate models evaluate unstart detection and new-onset prediction within a 0.5-s future interval. Seven Mach-2.9 runs, each with one annotated event, cover downstream throttling, throat regulation and splitter adjustment; three are held out with fixed weights and source-derived settings. Across three training seeds, full-sensor detection achieves run-wise mean window $F_1$ scores of 0.834--0.964. Eight-sensor aggregation reduces detection-score variation across sensor subsets in both splitter runs while retaining mean scores close to those of the full sets. Regional detection favors cowl and high-speed-channel measurements during splitter adjustment and compression-side measurements during downstream throttling, consistent with their respective pressure responses. Complete-record evaluation shows that configurations with comparable window scores can produce different alarm timing and persistence. These findings connect sensor count and regional coverage to pressure discrimination and sequential alarm behavior, providing experimental guidance for configuring inlet monitoring systems.
\end{abstract}

\end{frontmatter}

\section{Introduction}

Supersonic inlets capture and compress the air supplied to aeroengines. Maintaining a started flow is essential to matching the inlet air supply to engine demand, particularly when throat regulation or flow splitting changes the internal compression and flow distribution \cite{jin2023,chen2024,zhang2026restart}. Unstart disrupts this process through shock displacement and flow separation, producing changes in wall pressure and inlet operating state. Pressure measurements therefore provide a practical basis for identifying unstart and assessing whether an approaching transition can be detected early enough to support an operational response.

The pressure signature of this transition depends on both the operating path and the measurement location. Jin et al.\ \cite{jin2023} observed unstart--restart hysteresis during throat regulation and related its suppression to changes in the entrance separation-induced wave system. Chen et al.\ \cite{chen2024} identified different oscillatory stages during splitter-driven mode transition under unthrottled and throttled conditions. For low-speed-duct restart, Zhang et al.\ \cite{zhang2026restart} linked splitter motion to changes in local pressure oscillations and the timing of buzz. Angle-of-attack studies and time-resolved restart measurements further show that the evolution toward and away from unstart follows condition-dependent dynamics \cite{yang2025,johnson2026}. Spatially resolved pressure measurements reveal corresponding differences within the inlet/isolator \cite{bustard2024,bustard2026}. A pressure monitor must consequently relate local observations to the operating state without assuming that all regions exhibit the same response.

Time--frequency analysis, pressure-indicator fusion and learned pressure representations provide several routes to this monitoring problem \cite{trapier2007,chang2014,wang2022,wang2023}. Multiscale attention methods improve the representation of pressure histories, while operating-margin estimation and sparse-sensor recognition address the choice of prediction target and measurement set \cite{wang2025mmca,min2024,yang2026sparse}. For a monitor used with changing sensor configurations, two questions remain connected: how the available measurements affect state discrimination, and whether the resulting score trajectory produces a timely new alarm. A sustained positive output may cover the unstarted interval even when the alarm has been active well before onset. Window classification and continuous alarm histories are therefore both needed to assess the monitoring response.

This study investigates these questions using InletScope, a shared single-sensor model with multiscale pressure encoding and learned scale fusion. Arithmetic averaging of sensor probabilities allows the measurement set to change without retraining. Separate models perform detection and predict a new unstart onset within a 0.5-s future interval. The prediction interval defines the warning task, whereas the achieved lead is measured from a confirmed alarm declaration to the annotated onset. Model weights, source-derived normalization and validation-selected thresholds remain fixed during held-out evaluation.

Seven Mach-2.9 runs covering downstream throttling, throat regulation and splitter adjustment provide the experimental basis. Run-separated baseline comparisons and scale ablations assess the pressure representation, while sensor-count and regional comparisons isolate the effect of measurement composition at inference. Complete-record evaluation connects these comparisons to false alarms and declaration timing. The resulting contribution is a configurable pressure-monitoring method together with experimental evidence on sensing choices: eight-sensor aggregation reduces subset-dependent detection variation in the splitter runs, regional rankings change with operating condition, and continuous alarm histories relate pressure-classification responses to alarm activation and persistence.

\section{Related Work}

\subsection{Pressure-based inlet unstart detection and warning}

Pressure-based inlet monitoring relies on features that distinguish the observed flow states. Time--frequency analysis captures the changing oscillatory content of inlet buzz \cite{trapier2007}, and information-fusion methods combine pressure-derived indicators for backpressure-unstart detection \cite{chang2014}. Numerical studies also distinguish hard-unstart regimes with different geometric responses \cite{wang2026flow}. These results motivate representations that account for both temporal structure and operating condition, rather than treating every unstart record as an identical pressure transition.

Learning-based prediction introduces a second design choice: the definition of the state or event to be predicted. Wang et al.\ \cite{wang2022} formulate online unstart prediction through pressure-pattern recognition. Their subsequent approach combines water-cycle-optimized penalized-contrast change-point detection with single- and multichannel wavelet-packet CNNs, using the identified change points to partition flow states for training \cite{wang2023}. Such labels connect a model output to a pressure-pattern category. An event-relative target instead asks whether a new annotated onset follows a normal observation window within a specified interval. The distinction affects the meaning of a positive prediction and the reference time used to evaluate warning.

Operating-margin estimation offers another formulation. Min et al.\ \cite{min2024} use critical backpressure ratios to define unstart and warning states, and estimate backpressure with a one-dimensional CNN supplied by an optimized set of pressure measurements. Their framework combines numerical pressure data, mode-dependent models and wind-tunnel validation. It measures proximity to an operating boundary, whereas future-onset prediction measures the occurrence of a transition within a chosen time interval. The two formulations address different operational questions: how close the inlet is to a critical condition, and when a new event is expected.

The information available to either formulation depends on the measurement set. Yang et al.\ \cite{yang2026sparse} identify operating modes through dimensionality reduction, clustering and physical interpretation, then apply ReliefF sensor selection and gradient boosting across Mach numbers and sideboard configurations. Pressure-based flow-field reconstruction provides a complementary spatial description: Zhong et al.\ \cite{zhong2024} compare pressure-distribution and operating-condition inputs for reconstructing numerical inlet fields. These studies connect sensor placement to the information retained about the flow. For event monitoring, the corresponding requirement is to establish how selecting a different set of local pressure histories changes the classification and alarm sequence.

\subsection{Multiscale temporal learning and sensor-configuration transfer}

Multiscale representations describe pressure fluctuations at more than one temporal resolution. MMCANet combines multilevel wavelet-packet decomposition with attention-based fusion of multiscale time--frequency features for inlet unstart prediction \cite{wang2025mmca}. This is a direct precedent for learned weighting of scale-dependent pressure information. In time-series forecasting, TimeMixer combines decomposed seasonal and trend information across sampling scales \cite{wang2024timemixer}. InletScope uses a different scale construction: several patch lengths partition the same causal observation interval at a common sampling rate. The comparison with single-scale encoding and uniform fusion therefore examines temporal resolution and learned scale weighting within one pressure-classification architecture.

Spatial information and channel sharing provide different ways to handle a sensor array. MDEFNet combines temporal--spectral pressure features with a spatial encoder and evaluates recognition across sensor configurations, including a purged block-level split \cite{wang2026mdef}. WEMoE uses wavelet features and sensor-index-informed expert routing within known arrays, with adapter-based fine-tuning for condition changes \cite{wang2026wemoe}. These designs incorporate measurement identity into the learned representation or routing. A shared pressure-only encoder instead applies the same mapping to each available channel, leaving sensor composition to a subsequent aggregation step.

Channel-independent temporal learning has been studied in PatchTST, which shares patch embeddings and Transformer parameters across series \cite{nie2023}. iTransformer models interactions between variable tokens and evaluates forecasting on variates excluded from training \cite{liu2024itransformer}. The relevant design choice for inlet monitoring is whether the model combines sensors during feature extraction or after each pressure record has been classified. InletScope uses temporal self-attention and Pre-LN blocks \cite{vaswani2017,xiong2020} within a shared single-channel encoder, followed by arithmetic probability averaging. This produces an order-invariant output for a nonempty measurement set without introducing a learned sensor-fusion module.

Compatibility with a variable input set does not imply identical performance for every configuration. Each added or removed sensor changes the pressure histories contributing to the system score. Sensor-count, regional and withheld-tap tests therefore evaluate distinct aspects of the same fixed model: sensitivity to measurement number, dependence on the observed region, and response to taps excluded from fitting. In a computational control setting, the study by Mondal and Jagtap \cite{mondal2026} also examines changed backpressure, Reynolds number and sensor configurations using reinforcement learning. The present work concerns passive monitoring of experimental records, with sensor changes evaluated through both classification scores and alarm declarations.

\subsection{Event-oriented evaluation under distribution changes}

Window-level and event-level measures describe different aspects of a continuous monitor. Precision--recall analysis evaluates discrimination among labeled examples \cite{davis2006}, but the same classification outcomes can produce different alarm behavior when their temporal order changes. Sarfraz et al.\ \cite{sarfraz2024} discuss evaluation consistency and baseline selection in time-series anomaly detection, and PATE accounts for proximity between predicted and annotated anomaly intervals \cite{ghorbani2024}. For inlet monitoring, these considerations motivate reporting false-alarm episodes and declaration timing alongside window scores.

A warning declaration has a temporal requirement beyond overlap with a positive interval. An alarm that rises within the prediction interval supplies a new indication of an approaching onset. An alarm that was already active before that interval supplies no such transition, even if it remains active throughout unstart. Consecutive-output confirmation further changes the declaration time relative to the first threshold crossing. Event-level assessment must therefore follow the generated alarm sequence, including its normal-state activation and reset behavior.

Changing operating conditions also distinguishes direct transfer from adaptation. Kim et al.\ \cite{kim2024} update an anomaly detector at test time using trend estimation and self-supervision, while WEMoE uses adapter-based fine-tuning for new conditions \cite{wang2026wemoe}. Fixed-weight evaluation instead measures how a source-trained mapping responds without target-domain updates. Because probability calibration and discrimination are distinct properties \cite{guo2017}, the sensor-averaged probability in this study is used as a decision score with a source-validation threshold. Keeping preprocessing and decision settings unchanged makes configuration-dependent score shifts and alarm transitions observable under a common operating rule.

\section{Experimental data}

\subsection{Operating conditions and measurement layouts}

The experiments were conducted at a freestream Mach number of 2.9. Downstream-throttling runs use an internal contraction ratio of 1.50 at incidences of $0^\circ$, $+1^\circ$ and $+2^\circ$, with freestream static pressures of 3248.7, 3248.7 and 3245.8~Pa, respectively. The throat-regulation speeds are 5 and 10~mm\,s$^{-1}$ for trials A and B. Trial A has $0^\circ$ incidence and a freestream static pressure of 3237.4~Pa. The splitter-adjustment runs use speeds of 2 and 10~mm\,s$^{-1}$, both at $0^\circ$ incidence, with static pressures of 4300 and 4501~Pa.

For downstream throttling and throat regulation, taps C1--C13 and R1--R14 cover the cowl and compression surfaces, respectively. For splitter adjustment, A1--A7, B1--B7 and C1--C5 cover the cowl side, high-speed channel and compression side. Sensor identities combine the measurement layout and tap designation, and spatial comparisons use these physical regions.

\subsection{Pressure acquisition and data preparation}

Pressure is reported in kPa and time in seconds. Measurements were sampled at 20~kHz and subsampled to 2~kHz by retaining every tenth sample. Each processed pressure channel was matched to its original sensor record over the complete run, including the corresponding timestamps.

The processed records contain 27 matched pressure channels in each downstream run, 25 in each throat trial, and 19 and 17 in the two splitter runs. Unmatched, duplicate, padding and timestamp columns are excluded from the model input. Taps R12 and R13 are withheld from source fitting and evaluated separately on Downstream, $+2^\circ$, to examine transfer to previously unseen sensors. Table~\ref{tab:data} lists the sensor counts, run partition and class distribution.

\subsection{Event annotations and run partition}

The seven included runs each contain one annotated unstart onset: three runs supply training data, one supplies validation data and three are held out. A further $+3^\circ$ downstream record remains unstarted throughout and is excluded because it contains no observed onset. Training seeds and sensor subsets provide repeated evaluations of the same physical events. The held-out records had been inspected during earlier exploratory work and are excluded from fitting, normalization and threshold selection in the present run-separated protocol.

\begin{table}[!htbp]
\centering
\caption{Run partition and eligible windows. D and W denote detection and warning; counts are positive/eligible windows. Each run contains one annotated onset. Sensor counts exclude R12 and R13 from source fitting.}
\label{tab:data}
\small
\setlength{\tabcolsep}{4pt}
\begin{tabularx}{\textwidth}{@{}Xlrrr@{}}
\toprule
Condition & Partition & Sensors & D windows & W windows \\
\midrule
Downstream, $0^\circ$ & Training & 25 & 997/1832 & 50/785 \\
Downstream, $+1^\circ$ & Training & 25 & 1365/1849 & 50/484 \\
Throat, trial A & Training & 25 & 365/1039 & 50/624 \\
\addlinespace[3pt]
Throat, trial B & Validation & 25 & 210/775 & 50/515 \\
\addlinespace[3pt]
Downstream, $+2^\circ$ & Source test & 27 & 1345/1799 & 50/454 \\
Splitter, 2~mm\,s$^{-1}$ & Layout test & 19 & 287/804 & 50/467 \\
Splitter, 10~mm\,s$^{-1}$ & Layout test & 17 & 138/257 & 50/69 \\
\bottomrule
\end{tabularx}
\end{table}

\begin{figure}[!htbp]
\centering
\includegraphics[width=\textwidth]{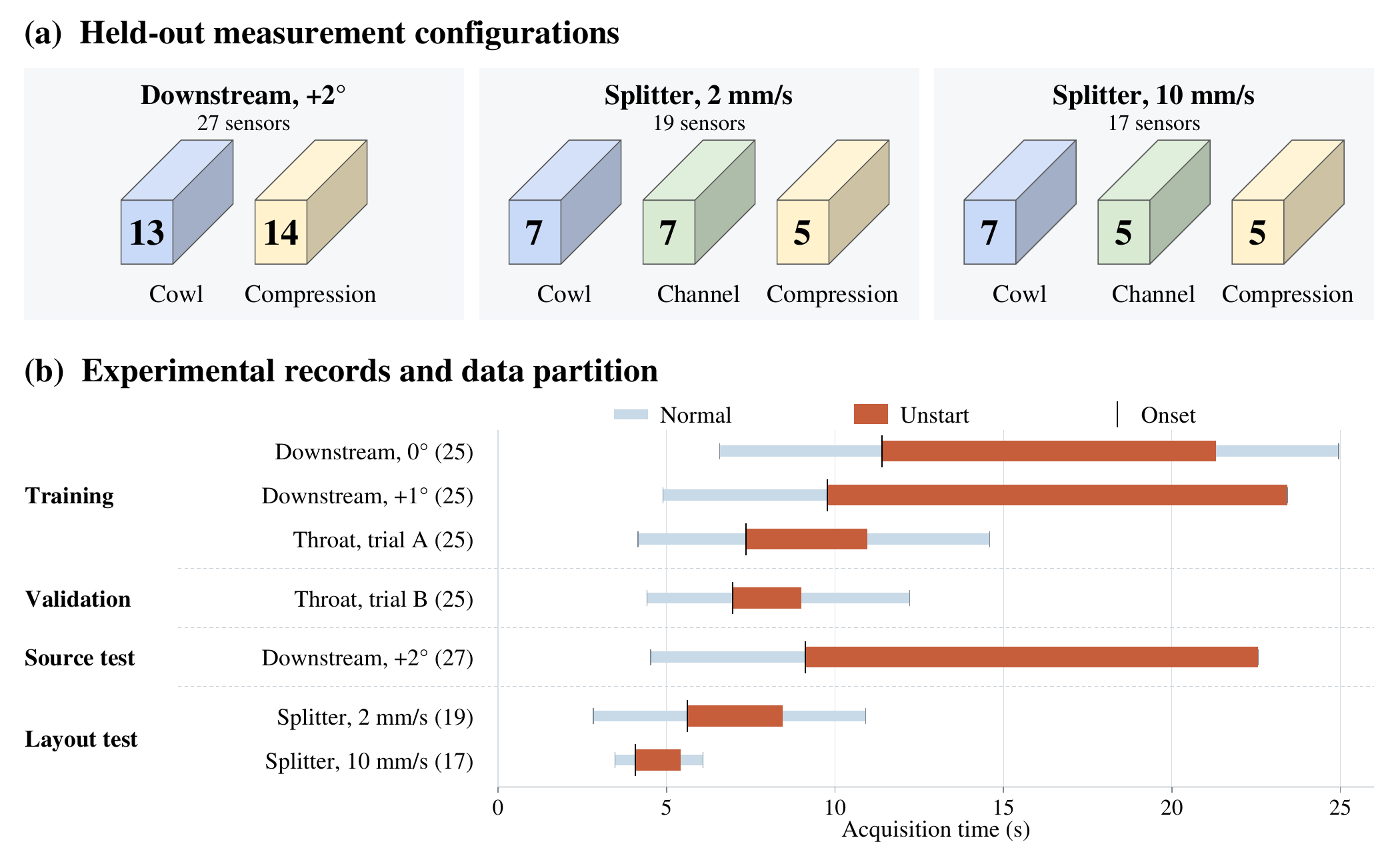}
\caption{Measurement configurations and experimental records. (a) Schematic sensor groupings in the three held-out runs, with regional counts on the block faces. Channel denotes the high-speed channel, and Compression the compression side. (b) Complete records on their original acquisition-time axes, grouped by data partition. Thin blue segments and thicker orange bands denote normal and unstart intervals, respectively; black ticks mark unstart onset. Dashed lines separate data partitions. Parentheses beside each operating condition give the included sensor count.}
\label{fig:data-map}
\end{figure}

Unstart and recovery times were taken from the annotations accompanying the experimental records. The downstream runs at $0^\circ$, $+1^\circ$ and $+2^\circ$ have onsets at 11.4, 9.7725 and 9.1187~s, respectively. The $0^\circ$ run recovers at 21.3172~s; the other two remain unstarted through the recorded endpoint. Throat trials A and B are unstarted over $[7.3622,10.955)$~s and $[6.9623,9.0067)$~s. The splitter intervals are $[5.62,8.44)$~s at 2~mm\,s$^{-1}$ and $[4.08,5.415)$~s at 10~mm\,s$^{-1}$. Each interval includes its onset and excludes its finite recovery endpoint, where present. Detection delays and warning lead times are measured relative to the annotated onsets. Figure~\ref{fig:data-map} summarizes the sensor regions and complete records.

\section{Shared-sensor monitoring method}

InletScope separates three operations: learning temporal pressure features, combining the available measurements and issuing sequential alarms. The learned component maps a single pressure window to a class probability. A parameter-free mean then maps any nonempty sensor set to one system score, and a common confirmation rule converts successive scores into declarations. This separation makes sensor configuration an input choice while preserving the task-specific temporal model.

\subsection{Causal detection and warning targets}
\label{sec:targets}

For run $r$, sensor $s$ and output index $j$, a window $\mathbf{x}_{r,s,j}\in\mathbb{R}^{100}$ contains the 100 pressure samples ending at output time $t_{r,j}$. The next output is 20 samples later, corresponding to a nominal 50-ms observation (49.5~ms between the first and last sample) and a 10-ms output stride. Model features comprise pressure amplitudes; timestamps locate each window in the record. The detection target is
\begin{equation}
y^D_{r,j}=\mathbb{I}\!\left[\text{any sample in }\mathbf{x}_{r,s,j}\text{ is annotated unstarted}\right].
\end{equation}
This target is identical across sensors of one run. For warning, only completely normal observation windows with a fully available future 0.5-s interval are eligible. If $u_r$ is the newly annotated onset, then
\begin{equation}
y^W_{r,j}=\mathbb{I}\!\left[t_{r,j}<u_r\leq t_{r,j}+0.5\,\mathrm{s}\right].
\end{equation}
The interval begins immediately after the observed window, with zero gap. Windows containing an unstarted sample or lacking a complete future interval are excluded from warning training and window scoring. During continuous inference, every incoming window is evaluated regardless of its annotated state.

\subsection{Source-fitted pressure normalization}

Let $v_{r,s,n}$ denote sample $n$ of sensor $s$ in source-training run $r$, with $T_r$ samples and $S_r$ included sensors. A single location and scale are shared by both tasks and all sensors:
\begin{align}
\mu&=\frac{1}{|\mathcal R_{\rm tr}|}\sum_{r\in\mathcal R_{\rm tr}}
 \frac{1}{S_rT_r}\sum_{s=1}^{S_r}\sum_{n=1}^{T_r}v_{r,s,n},\\
m_2&=\frac{1}{|\mathcal R_{\rm tr}|}\sum_{r\in\mathcal R_{\rm tr}}
 \frac{1}{S_rT_r}\sum_{s=1}^{S_r}\sum_{n=1}^{T_r}v_{r,s,n}^{\,2},
\qquad \sigma=\sqrt{\max(m_2-\mu^2,10^{-12})}.
\label{eq:normalization}
\end{align}
Each training run receives equal weight, as does each sensor within a run. These moments use the complete training pressure records before window sampling. The resulting $\mu=16.5170623$~kPa and $\sigma=12.2757170$~kPa transform every window as $\widetilde{\mathbf x}=(\mathbf x-\mu)/\sigma$. The same constants apply unchanged to all validation and test inputs.

\subsection{Multiscale single-sensor encoder}

Patch length controls the granularity of temporal representation. Short patches produce more tokens describing local signal segments, while longer patches group samples before attention. Parallel encoders therefore provide several resolutions of the same observed interval, with scale attention determining their contributions to the single-sensor output.

InletScope divides each normalized single-sensor window into non-overlapping patches of lengths $\ell\in\{4,10,20\}$, giving $N_\ell=100/\ell$ tokens. The corresponding patch durations are 2, 5 and 10~ms at 2~kHz. For a patch matrix $P_\ell\in\mathbb R^{N_\ell\times\ell}$, the initial tokens are $Z_\ell^{(0)}=P_\ell W_\ell+\mathbf 1b_\ell^\top+{\rm PE}_\ell$, where $W_\ell\in\mathbb R^{\ell\times64}$ is scale-specific. Fixed sinusoidal positional encoding uses $\operatorname{PE}(n,2i)=\sin(n/10000^{2i/64})$ and $\operatorname{PE}(n,2i+1)=\cos(n/10000^{2i/64})$, with zero-based token index $n$ and $i=0,\ldots,31$ \cite{vaswani2017}. All 100 samples are retained at every scale.

Multiscale patch division follows the temporal-resolution principle explored by Pathformer \cite{chen2024pathformer}. InletScope evaluates all three scales with standard self-attention and dense scale weighting, whereas Pathformer combines intra-/inter-patch attention with adaptive pathway routing.

Each encoder contains two separately parameterized layers with four attention heads, feature width 64 and feed-forward width 128. For a normalized token matrix $U$, each head forms affine projections $Q_i,K_i,V_i\in\mathbb R^{N_\ell\times16}$ and computes
\begin{align}
{\rm head}_i&={\rm Dropout}\!\left[{\rm softmax}
 \left(\frac{Q_iK_i^\top}{\sqrt{16}}\right)\right]V_i,\\
{\rm MHA}(U)&={\rm Concat}({\rm head}_1,\ldots,{\rm head}_4)W^O+b^O.
\end{align}
Softmax acts over key tokens. Layer normalization precedes attention and the feed-forward sublayer (Pre-LN), with affine parameters and $\epsilon=10^{-5}$ \cite{xiong2020}. Within a layer, the residual updates are
\begin{align}
\widetilde{Z}&=Z+\operatorname{Dropout}\!\left[\operatorname{MHA}(\operatorname{LN}(Z))\right],\\
Z^{+}&=\widetilde{Z}+\operatorname{Dropout}\!\left[\operatorname{FFN}(\operatorname{LN}(\widetilde{Z}))\right].
\end{align}
The feed-forward sublayer uses a $64\!\to\!128\!\to\!64$ affine mapping with ReLU and intermediate dropout. Attention-probability, feed-forward and residual-output dropout all use rate 0.1 and are disabled at inference. Token pooling follows the final residual block directly. Unmasked attention operates within the observed window, preserving causality with respect to its endpoint.

\begin{figure}[!htbp]
\centering
\includegraphics[width=\textwidth]{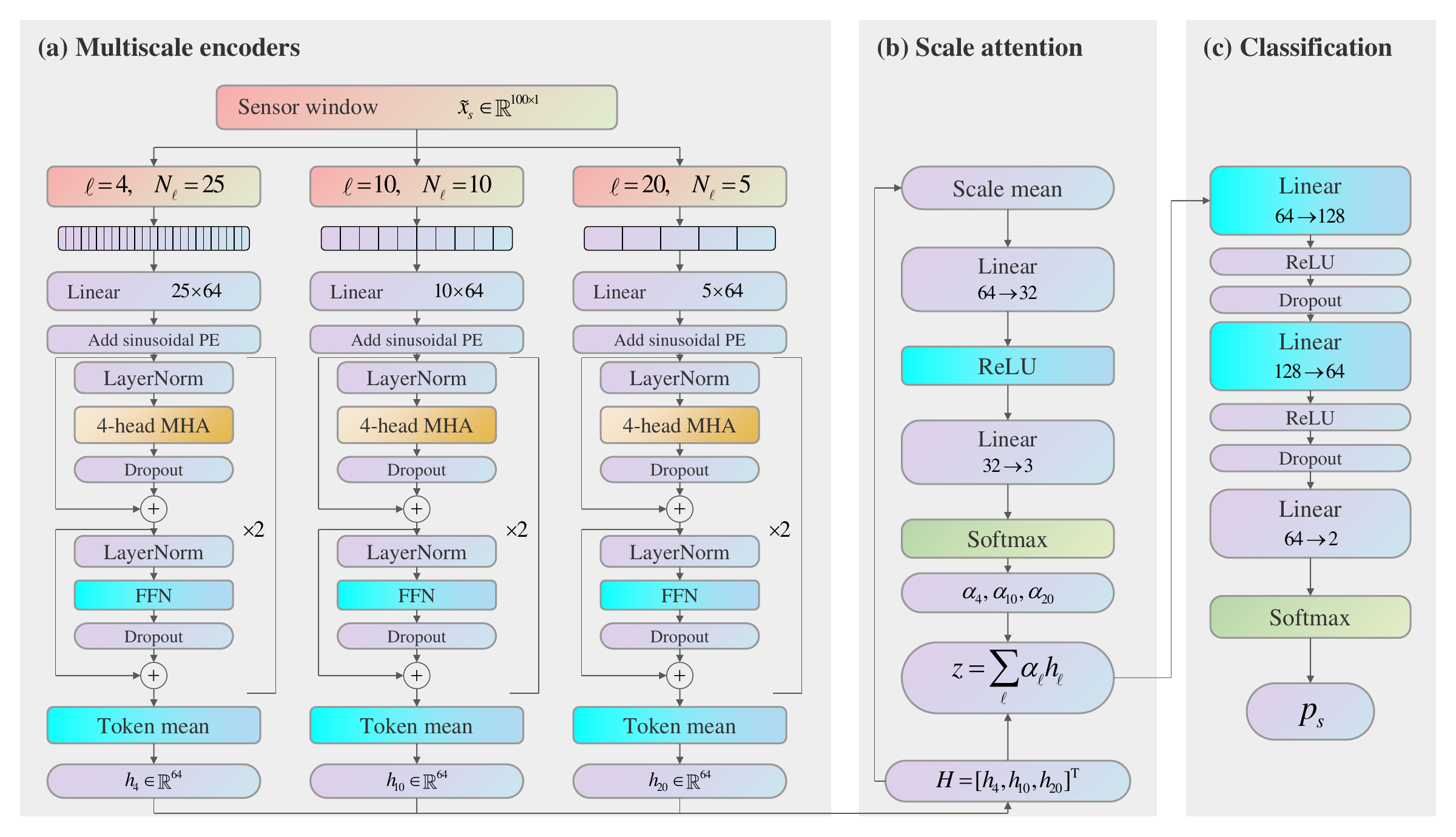}
\caption{InletScope single-sensor architecture. (a) Patch lengths 4, 10 and 20 yield 25, 10 and 5 tokens. Linear projections and sinusoidal positional encoding (PE) precede independent two-layer Pre-LN encoders and token-mean pooling. (b) The mean of stacked scale features $H$ generates weights $\alpha$ for their weighted sum. (c) The classifier returns two logits before softmax. Tensor sizes omit the batch dimension. Graphic elements are adapted from ML Visuals (MIT license).}
\label{fig:architecture}
\end{figure}

Token-wise mean pooling produces a 64-dimensional vector $h_\ell$ at each scale. Stacking these vectors gives $H\in\mathbb{R}^{3\times64}$. The scale gate first averages the three vectors and then computes normalized weights:
\begin{equation}
g=\frac{1}{3}\sum_\ell h_\ell,\qquad
\alpha=\operatorname{softmax}\!\left[W_2\operatorname{ReLU}(W_1g+b_1)+b_2\right],
\end{equation}
where $W_1\in\mathbb{R}^{32\times64}$ and $W_2\in\mathbb{R}^{3\times32}$. The fused representation is $z=\sum_\ell\alpha_\ell h_\ell$. A $64\!\to\!128\!\to\!64\!\to\!2$ classifier produces logits, with ReLU and dropout after the first two linear layers. Softmax converts the logits to a single-sensor positive-class probability. Detection and warning have separate fitted parameters; within either task, the same complete model is applied to every sensor. Figures~\ref{fig:method} and \ref{fig:architecture} show the single-sensor model and parameter-free aggregation.

\subsection{Sensor aggregation and sequential alarms}

For a nonempty set of available sensors $\mathcal S_r$, the system probability is
\begin{equation}
q_{r,j}(\mathcal S_r)=\frac{1}{|\mathcal S_r|}\sum_{s\in\mathcal S_r}
\mathrm{softmax}\!\left[f_{\theta}(\widetilde{\mathbf{x}}_{r,s,j})\right]_{1}.
\label{eq:mean}
\end{equation}
The positive class is unstart for detection and future onset for warning. The arithmetic mean is invariant to sensor ordering and reduces to single-sensor inference for a singleton. Only available sensors contribute to the numerator and denominator. An empty set returns no observation and no risk value.

\begin{figure}[!htbp]
\centering
\includegraphics[width=\textwidth]{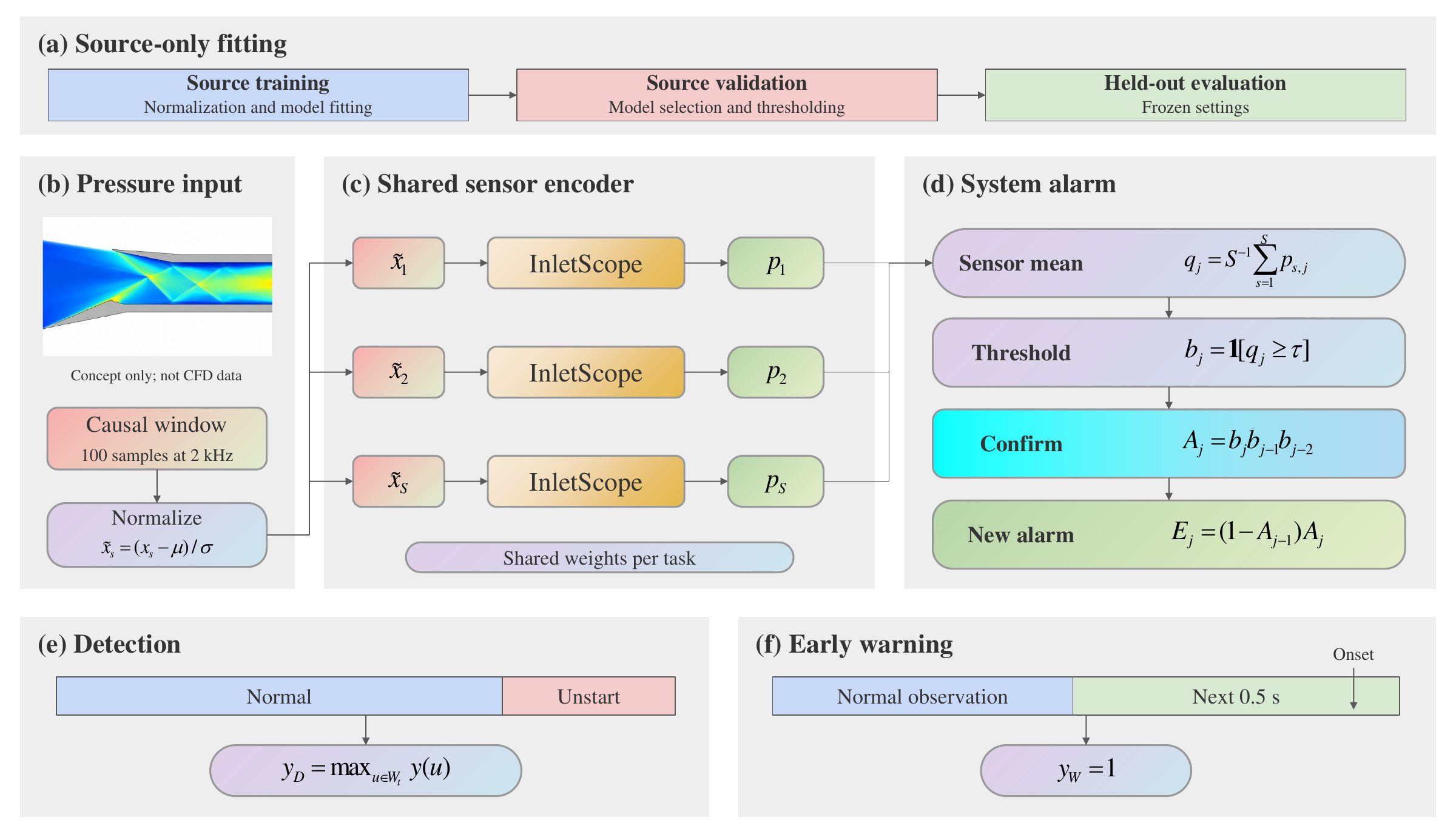}
\caption{Shared single-sensor monitoring. (a) Source training and held-out evaluation with fixed settings. (b) Causal pressure windows with source-fitted normalization. (c) Shared single-sensor inference, with separate models for detection and warning. (d) Probability averaging over $S$ sensors, thresholding, three-output confirmation $A_j$ and new-alarm edge $E_j$. (e) A positive detection window contains unstart; 100 samples at 2~kHz span 49.5~ms. (f) A positive warning window is normal and precedes a new onset (arrow) within 0.5~s. Interval widths are schematic. The inlet inset is an AI-generated conceptual illustration, not CFD data. Graphic elements are adapted from \href{https://github.com/dair-ai/ml-visuals}{ML Visuals} (MIT license).}
\label{fig:method}
\end{figure}

For a fitted model, one validation-selected threshold $\tau$ is shared across test runs and sensor configurations. With $b_j=\mathbb I[q_{r,j}\geq\tau]$, the confirmed alarm and its declaration edge are
\begin{equation}
A_j=\prod_{i=0}^{2}b_{j-i},\qquad E_j=A_j(1-A_{j-1}),
\label{eq:alarm}
\end{equation}
where unavailable pre-record outputs and $A_0$ are zero. Three-output confirmation adds 20~ms between the first above-threshold output and declaration. A below-threshold output immediately clears the alarm and resets confirmation; each new run starts with a cleared state. Direct thresholding uses $A_j=b_j$. Both rules operate on the same probability sequence and threshold, with declarations timestamped when their respective conditions are met. Figure~\ref{fig:method} shows the data flow and target intervals.

\section{Evaluation protocol}
\subsection{Comparison design and baselines}

The comparisons follow the three components of the monitor. Temporal-convolution and statistical-feature baselines assess pressure representation, sensor subsets assess measurement composition, and direct versus confirmed alarms assess temporal decision rules. InletScope, a temporal convolutional network (TCN) and a statistical-feature random forest (RF) use the same run partition, eligible windows, sensor sets and evaluation procedure. Every method is trained on single-sensor inputs and uses the probability mean in Eq.~\ref{eq:mean} at system level. Evaluation covers held-out operating conditions and layouts, with saved probability sequences reused across sensor configurations and alarm rules.

The TCN is a compact single-convolution variant of the causal residual architecture of Bai et al.\ \cite{bai2018tcn}. Its three blocks have output width 64, kernel width five and dilations $d=1,2,4$. Each block implements $\operatorname{ReLU}[{\rm Conv}_{5,d}({\rm PadLeft}_{4d}(X))+{\rm Conv}_{1}(X)]$, with stride one and a learned $1\times1$ skip projection. The convolutional stack has a 29-sample receptive field per output position; mean pooling over all 100 positions provides the window representation. Its $64\!\to\!128\!\to\!64\!\to\!2$ head applies ReLU after the first two affine layers and dropout 0.1 after the first ReLU. The convolutional blocks use neither normalization nor dropout.

RF \cite{breiman2001} uses eight features computed from the normalized window: mean, population standard deviation, minimum, maximum, median, final value, final-minus-initial value and mean absolute adjacent difference. The forest contains 300 bootstrap-sampled trees, with Gini splitting, $\lfloor\sqrt{8}\rfloor=2$ candidate features per split, minimum leaf size five, minimum split size two and unrestricted depth. Class weighting is set to balanced; the balanced sampled training set gives equal class weights. Four CPU workers are used. Probabilities are the mean of tree-level class probabilities.

Two InletScope variants isolate architectural choices. The single-scale model retains patch length ten, one two-layer encoder, the one-output scale gate and the same classifier. The uniform-fusion model retains all three encoders but replaces the learned gate with fixed weights $1/3$. Each variant is fitted independently with the same task-specific protocol. The five models are evaluated on both tasks with three training seeds, giving 30 fits in total.

\subsection{Sampling, optimization and model selection}

For each task, model and seed (17, 29 and 43), an epoch samples 4000 eligible windows with replacement from each class in each of the three source runs. For every sampled window, a sensor is drawn independently with probability $1/S_r$ from the included sensors of that run. The resulting 24,000 examples are shuffled and divided into 94 nearly equal minibatches of 255 or 256 examples. This scheme balances run--class contributions and assigns equal selection probability to sensors within each run. The draw is refreshed each neural epoch; RF uses one 24,000-example draw. Validation and test windows are evaluated exhaustively.

Neural models minimize unweighted two-class cross-entropy on sampled single-sensor examples,
\begin{equation}
\mathcal L(\theta)=-\frac{1}{B}\sum_{i=1}^{B}
\log {\rm softmax}\!\left[f_\theta(\widetilde{\mathbf x}_i)\right]_{y_i}.
\end{equation}
Training uses hard class labels and unaugmented single-sensor windows. AdamW \cite{loshchilov2019} uses a constant learning rate of $10^{-3}$, weight decay $10^{-4}$, $\beta_1=0.9$, $\beta_2=0.999$ and $\epsilon=10^{-8}$. Global gradient norm is clipped to one. All neural models train from initialization in float32. The InletScope encoders are cloned from one initialized PyTorch encoder layer; corresponding initial values coincide, and each layer's parameters are optimized independently. Python, NumPy and PyTorch generators use the stated seed; cuDNN deterministic mode is enabled and its benchmark mode is disabled.

After each epoch, positive-class probabilities are averaged over the 25 validation sensors, and average precision (AP) is computed on that task's eligible validation windows. A checkpoint is replaced when AP increases by more than $10^{-5}$. Training stops after eight epochs without such an improvement or at epoch 40, then restores the selected checkpoint. For each fitted model, $\tau$ maximizes validation $F_1$ over $\{0.01,0.02,\ldots,0.99\}$, with ties resolved toward the higher threshold. This threshold is selected before alarm confirmation and retained for every test sensor set. Architecture and optimizer settings are fixed across the comparison.

Neural fitting uses an NVIDIA GeForce RTX 3060 and RF fitting uses an Intel Core i7-14700 CPU. The software environment is Windows, Python 3.11.14, PyTorch 2.6.0+cu124, NumPy 2.1.2 and scikit-learn 1.7.2. Across the 24 neural fits, training terminates after 9--16 epochs and selects checkpoints from epochs 1--8. For InletScope seeds 17, 29 and 43, the selected detection epochs are 2, 2 and 3, with thresholds 0.27, 0.25 and 0.22; the warning epochs are 2, 8 and 2, with thresholds 0.41, 0.41 and 0.39. Table~\ref{tab:protocol} summarizes the common settings.

\begin{table}[t]
\centering
\caption{Training and inference settings shared by the comparison methods. All selection uses the source-training and validation runs.}
\label{tab:protocol}
\small
\renewcommand{\arraystretch}{1.12}
\begin{tabularx}{\textwidth}{@{}lX@{}}
\toprule
Component & Fixed setting \\
\midrule
Pressure input & 2 kHz; 100 points; stride 20 \\
Warning horizon & 0.5 s; gap 0; complete future \\
Source partition & Three training runs and one validation run (Table~\ref{tab:data}) \\
Seeds & 17, 29, 43 \\
Neural optimizer & AdamW; constant learning rate $10^{-3}$; weight decay $10^{-4}$; unweighted cross-entropy \\
Training schedule & Maximum batch 256; at most 40 epochs; patience 8; gradient norm clipped at 1 \\
Epoch sampling & 4000 windows per run and class, with replacement; independent uniform sensor draws \\
Checkpoint & Validation mean-probability AP; improvement $>10^{-5}$ \\
Threshold & Validation $F_1$, grid 0.01--0.99; highest threshold on ties \\
Alarm & Three successive outputs; reset on the first below-threshold output \\
Sensor fusion & arithmetic probability mean \\
\bottomrule
\end{tabularx}
\end{table}

\subsection{Sensor configurations}
Held-out sensor counts are 1, 2, 4, 8 and all included sensors. For each non-full count, up to 20 distinct subsets are drawn without replacement within a subset; subsets may overlap with one another. The count is $\min[20,\binom{S_r}{k}]$, giving 17 singleton configurations for the 10~mm\,s$^{-1}$ run. A deterministic generator derived from master seed 20261002, run identity and sensor count fixes these lists before evaluation; every method and training seed uses the same lists. Exact sensor IDs are provided in the method package.

Region groups follow the physical labels in the experiment description. Region-balanced coverage sets allocate four or eight sensors round-robin across the cowl and compression regions in the downstream layout, and across the cowl, high-speed-channel and compression regions in the splitter layouts. Within each region, selected indices are rounded from an evenly spaced sequence between the first and last included sensor; a single selection uses the rounded central index. This combines regional coverage with index-spaced sampling. R12 and R13 in Downstream, $+2^\circ$ provide a separate test of sensors excluded from source fitting. The remaining subsets assess configuration sensitivity.

\subsection{Window and event-level measures}
Window precision, recall, $F_1$ and AP use the positive-class definitions in Section~\ref{sec:targets} and the task's eligible windows. At threshold $\tau$, precision is $\mathrm{TP}/(\mathrm{TP}+\mathrm{FP})$, recall is $\mathrm{TP}/(\mathrm{TP}+\mathrm{FN})$ and $F_1=2\mathrm{TP}/(2\mathrm{TP}+\mathrm{FP}+\mathrm{FN})$; a zero denominator returns zero. AP is the non-interpolated precision--recall sum implemented by scikit-learn. Precision, recall and $F_1$ use thresholded probabilities before confirmation; AP uses continuous probabilities.

Event coverage means the detection alarm is active at any output time inside the annotated unstart interval $[u_r,v_r)$, where $v_r$ is the recovery time or $+\infty$ when recovery is not annotated. A new detection declaration requires $E_j=1$ in that interval. A valid warning requires $E_j=1$ in $[u_r-0.5\,{\rm s},u_r)$. The first qualifying edge defines detection delay $t_j-u_r$ or warning lead $u_r-t_j$. A pre-existing alarm can cover the event without supplying a new declaration; when no qualifying edge occurs, timing is undefined.

False-alarm episodes are connected components of the active alarm intersected with the normal evaluation intervals. For detection, those intervals are the annotated normal portions of the record. For warning, they exclude the 0.5-s pre-onset interval and the final incomplete future horizon, but include eligible post-recovery time. An alarm persisting across recovery therefore contributes a false episode in the normal segment. The rate is $60N_{\rm FA}/T_{\rm normal}$, with $T_{\rm normal}$ in seconds computed from raw timestamps under the same interval rules. Active-alarm fraction is $\sum_j A_j/J$ over the complete $J$-output record, including unstart and, for warning, censored tail outputs.

Metrics are reported separately for each held-out run. Tables use the arithmetic mean and sample standard deviation (denominator $3-1$) over the three training seeds. In sensor-count plots, subset quantiles are computed after averaging each fixed subset over seeds; seed ranges are computed after averaging each seed over subsets. These summaries describe fitting and configuration variability within each physical run.

A retrospective pressure comparison complements the regional scores. For each region, the median across per-sensor temporal medians summarizes the 0.5-s intervals immediately before and after the annotated onset.

\section{Results}

\subsection{Held-out detection and warning}

\begin{figure}[!htbp]
\centering
\includegraphics[width=\textwidth]{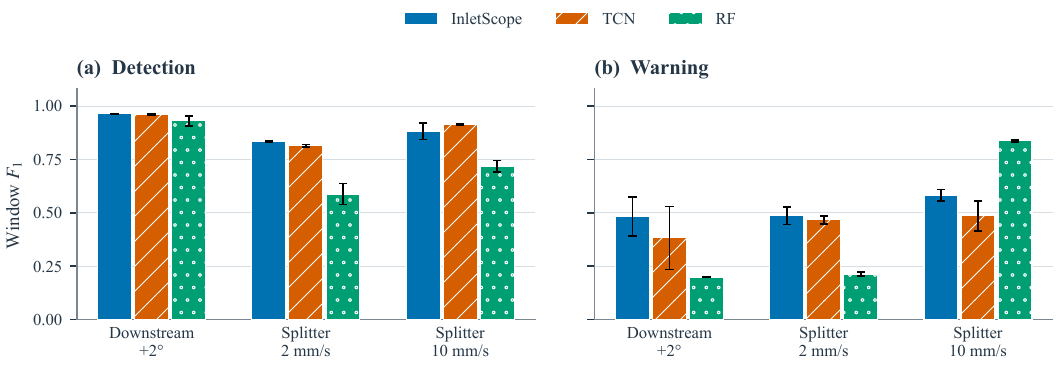}
\caption{Held-out window classification with all included sensors. Bars show mean $F_1$ and whiskers show sample standard deviation over seeds 17, 29 and 43. Color and hatching distinguish methods. Both tasks use source-validation thresholds.}
\label{fig:main}
\end{figure}

The shared model retains detection discrimination across the held-out operating conditions and layouts. InletScope achieves window $F_1$ values of $0.964\pm0.001$, $0.834\pm0.003$ and $0.882\pm0.039$ on Downstream, $+2^\circ$, Splitter, 2~mm\,s$^{-1}$ and Splitter, 10~mm\,s$^{-1}$, respectively (Fig.~\ref{fig:main}). Its mean detection $F_1$ exceeds RF on all three runs and TCN on the first two. TCN has the highest mean on the 10~mm\,s$^{-1}$ run.

The corresponding alarm histories distinguish state discrimination from onset declaration (Table~\ref{tab:main}). For every method and seed, including the ablations, the full-sensor detection alarm activates before onset and covers the event without a new declaration during unstart. Event coverage in these records is therefore produced by continued activation across the annotated onset.

\begin{table}[!htbp]
\centering
\caption{Full-sensor classification and confirmed alarms. $F_1$ is mean $\pm$ sample SD across three seeds. False episodes, false episodes per normal minute and active-alarm fraction are seed means. All detection alarms are already active before onset; the text reports valid warning leads by seed. False/min uses the observed normal exposure, which is 0.28~s for the 10~mm\,s$^{-1}$ warning case. Active denotes the fraction of the complete output record with an active alarm.}
\label{tab:main}
\small
\setlength{\tabcolsep}{5pt}
\begin{tabular*}{\textwidth}{@{\extracolsep{\fill}}lcrrr@{}}
\toprule
Model & $F_1\uparrow$ & False$\downarrow$ & False/min$\downarrow$ & Active \\
\midrule
\multicolumn{5}{l}{\textbf{Detection: downstream, $+2^\circ$}} \\
InletScope & $0.964\pm0.001$ & 1.7 & 21.8 & 0.80 \\
TCN & $0.962\pm0.003$ & 1.3 & 17.4 & 0.81 \\
RF & $0.931\pm0.024$ & 8.7 & 113.4 & 0.84 \\
\addlinespace[3pt]
\multicolumn{5}{l}{\textbf{Detection: splitter, 2~mm\,s$^{-1}$}} \\
InletScope & $0.834\pm0.003$ & 2.3 & 26.6 & 0.50 \\
TCN & $0.815\pm0.005$ & 2.0 & 22.8 & 0.52 \\
RF & $0.588\pm0.049$ & 17.0 & 193.8 & 0.81 \\
\addlinespace[3pt]
\multicolumn{5}{l}{\textbf{Detection: splitter, 10~mm\,s$^{-1}$}} \\
InletScope & $0.882\pm0.039$ & 2.0 & 93.9 & 0.67 \\
TCN & $0.914\pm0.003$ & 2.0 & 93.9 & 0.63 \\
RF & $0.719\pm0.027$ & 3.7 & 172.2 & 0.93 \\
\midrule
\multicolumn{5}{l}{\textbf{Warning: downstream, $+2^\circ$}} \\
InletScope & $0.484\pm0.091$ & 2.0 & 29.4 & 0.83 \\
TCN & $0.383\pm0.147$ & 5.3 & 78.3 & 0.88 \\
RF & $0.198\pm0.000$ & 1.0 & 14.7 & 1.00 \\
\addlinespace[3pt]
\multicolumn{5}{l}{\textbf{Warning: splitter, 2~mm\,s$^{-1}$}} \\
InletScope & $0.486\pm0.041$ & 2.7 & 37.5 & 0.55 \\
TCN & $0.468\pm0.019$ & 2.3 & 32.8 & 0.56 \\
RF & $0.213\pm0.009$ & 9.7 & 136.0 & 0.91 \\
\addlinespace[3pt]
\multicolumn{5}{l}{\textbf{Warning: splitter, 10~mm\,s$^{-1}$}} \\
InletScope & $0.582\pm0.028$ & 1.0 & 216.2 & 0.67 \\
TCN & $0.486\pm0.071$ & 1.0 & 216.2 & 0.66 \\
RF & $0.837\pm0.006$ & 2.0 & 432.4 & 0.96 \\
\bottomrule
\end{tabular*}
\end{table}

Warning $F_1$ for InletScope is $0.484\pm0.091$, $0.486\pm0.041$ and $0.582\pm0.028$ on the same runs. Valid full-sensor warning declarations are confined to the 10~mm\,s$^{-1}$ splitter event for all three methods. On this event, InletScope gives confirmed warning leads of 394, 204 and 204~ms for seeds 17, 29 and 43. TCN gives corresponding leads of 154, 154 and 404~ms. RF has the highest window $F_1$ on this run and gives a 304-ms warning for seed 43; seeds 17 and 29 produce no valid warning. InletScope produces one false-alarm episode per seed over approximately 0.28~s of eligible normal exposure, equivalent to 216 episodes per normal minute.

\subsection{Number and region of sensors}

\begin{figure}[!htbp]
\centering
\includegraphics[width=\textwidth]{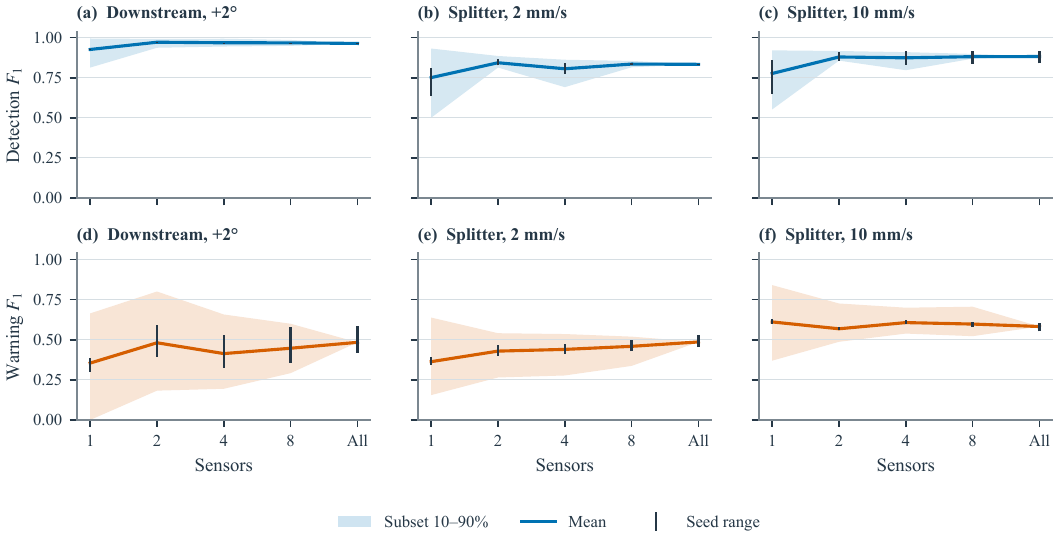}
\caption{InletScope sensitivity to sensor count. Lines show mean window $F_1$. Shading at non-full counts spans the 10th--90th percentiles of fixed-subset scores after averaging each subset across seeds. Vertical strokes show the range of the three seed means after averaging each seed across subsets. The count levels are categorical; All denotes 27, 19 and 17 sensors in the respective columns. Each non-full count uses 20 unique subsets, except the singleton sets in the two splitter runs, which enumerate all 19 and 17 sensors, respectively.}
\label{fig:sensors}
\end{figure}

Eight-sensor subsets reduce detection-score variation across sensor choices in both splitter runs (Fig.~\ref{fig:sensors}). The 10th--90th percentile interval of seed-averaged subset $F_1$ narrows from 0.501--0.932 with one sensor to 0.815--0.855 with eight sensors at 2~mm\,s$^{-1}$, and from 0.552--0.921 to 0.869--0.898 at 10~mm\,s$^{-1}$. Eight-sensor mean scores, 0.835 and 0.881, are close to the full-set scores of 0.834 and 0.882. The intermediate count levels are non-monotonic, and variation across training seeds remains visible. Some single- and two-sensor sets on Downstream, $+2^\circ$ produce new detection edges, and several fixed subsets yield valid warnings.

\begin{figure}[!htbp]
\centering
\includegraphics[width=\textwidth]{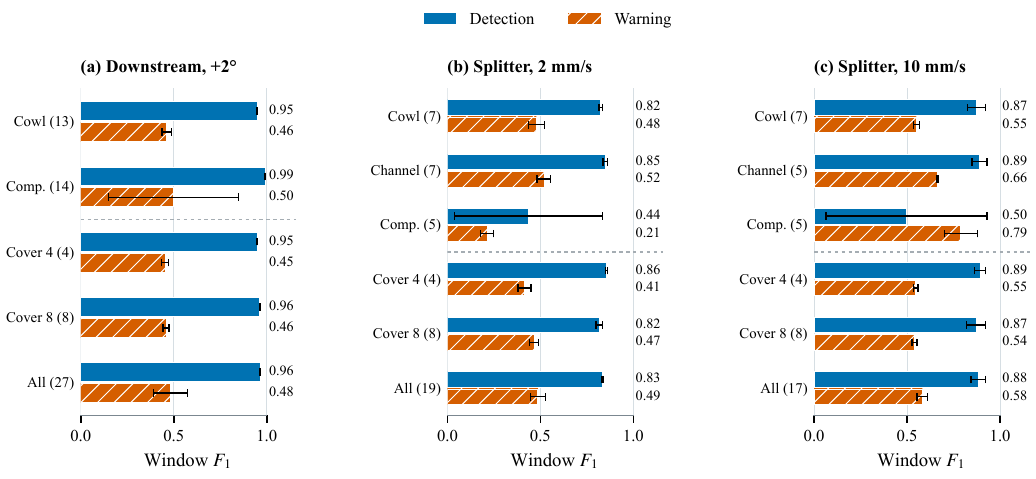}
\caption{InletScope performance by physical region and region-balanced sensor coverage. Bars and right-hand values show mean window $F_1$; whiskers show sample SD over seeds 17, 29 and 43. Parentheses give the sensor count in each configuration. The downstream layout has cowl and compression-side groups; the splitter layouts also have a high-speed-channel group. Dashed separators distinguish regional groups from coverage/all-sensor sets. Cover 4/8 denotes region-balanced, index-spaced selection. Both tasks use the same horizontal scale.}
\label{fig:regions}
\end{figure}

Regional detection rankings change with operating condition (Fig.~\ref{fig:regions}). At splitter speeds of 2 and 10~mm\,s$^{-1}$, the high-speed-channel group gives mean $F_1$ of 0.849 and 0.890, followed by the cowl group at 0.824 and 0.873; the compression group gives 0.436 and 0.497. On Downstream, $+2^\circ$, compression-side measurements give 0.993, exceeding the cowl-side value of 0.947.

The pressure histories show distinct regional responses across these conditions. In the downstream run, the pre-/post-onset pressure levels are 34.6/39.2~kPa at the cowl and 7.2/21.2~kPa on the compression side. At a splitter speed of 2~mm\,s$^{-1}$, the cowl and high-speed-channel levels rise from 26.4 to 31.2~kPa and from 25.7 to 29.4~kPa, respectively, while the compression-side level remains near 2.6~kPa. At 10~mm\,s$^{-1}$, the corresponding levels are 11.1/31.5, 10.3/29.7 and 5.6/2.6~kPa. These condition-specific pressure responses accompany the changes in regional detection performance.

\subsection{Complete alarm histories and fixed-rule trade-off}

\begin{figure}[!htbp]
\centering
\includegraphics[width=\textwidth]{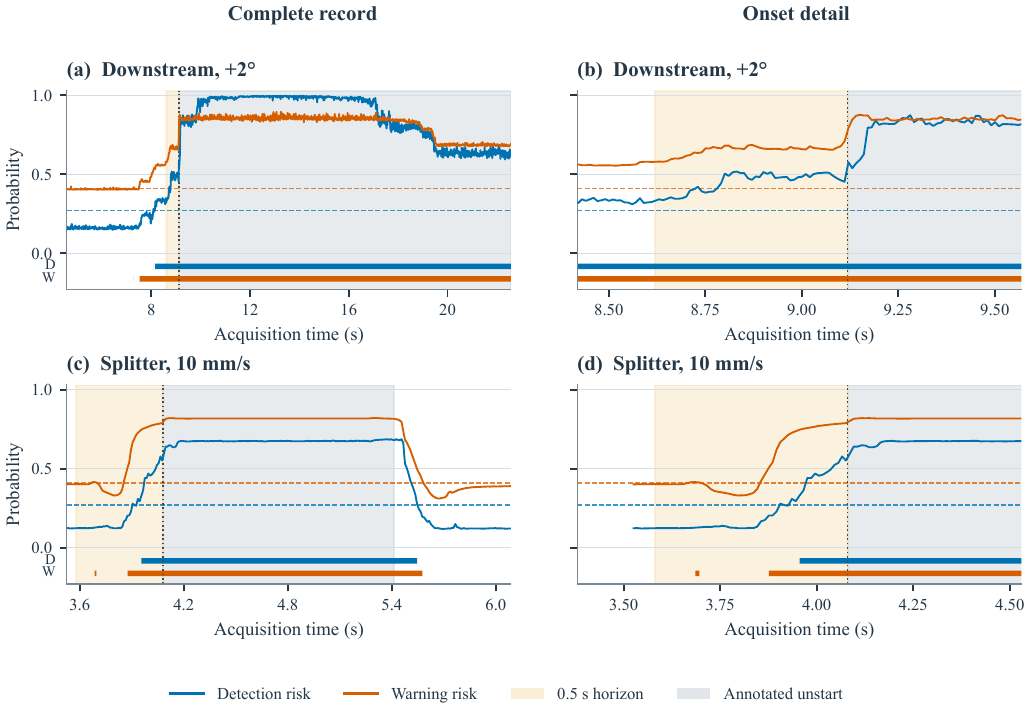}
\caption{Complete and onset-local histories for InletScope seed 17 using all sensors. Blue and orange traces are the respective detection and warning mean probabilities; dashed lines are their source-validation thresholds. Horizontal ribbons below zero mark active three-output alarms. Pale orange is the 0.5-s warning interval preceding onset; gray is annotated unstart. The event already lies inside an active detection alarm on both runs, whereas the Splitter, 10~mm\,s$^{-1}$ warning has a new edge inside the permitted interval.}
\label{fig:timelines}
\end{figure}

\begin{figure}[!htbp]
\centering
\includegraphics[width=\textwidth]{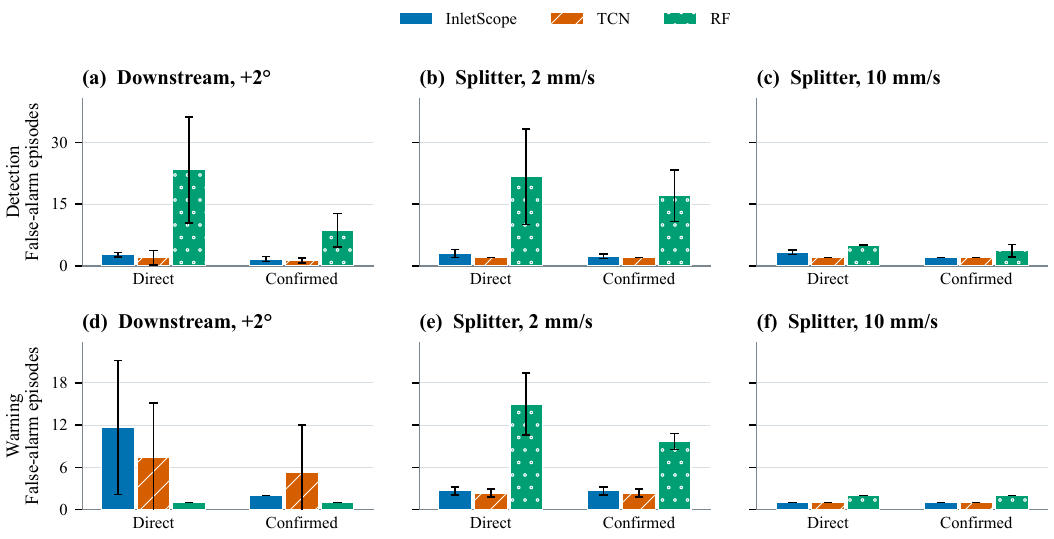}
\caption{False-alarm episodes under direct thresholding and three-output confirmation at the same source-validation threshold. Bars show means and whiskers sample SD over seeds 17, 29 and 43, using all sensors. Colors and hatching follow Fig.~\ref{fig:main}; vertical scales are shared within each task. Exposure-normalized rates are reported in Table~\ref{tab:main}.}
\label{fig:tradeoff}
\end{figure}

Complete alarm histories explain how positive classifications become persistent activation or new declarations (Fig.~\ref{fig:timelines}). For the predefined seed 17 on Downstream, $+2^\circ$, both task probabilities cross their thresholds before the event and remain elevated into unstart, producing no qualifying new declaration. On Splitter, 10~mm\,s$^{-1}$, the warning alarm rises inside the 0.5-s pre-onset interval and remains active into the event. The detection alarm rises before onset, and the full record shows continued alarm activity after recovery. Comparing direct and three-output rules at the same thresholds (Fig.~\ref{fig:tradeoff}) separates brief excursions from sustained responses: confirmation suppresses the former while the latter persist.

Regional warning scores show the same distinction. On Splitter, 10~mm\,s$^{-1}$, the compression group has higher mean warning $F_1$ (0.789) than the cowl (0.550) or high-speed-channel group (0.662), yet only its seed-29 fit produces a valid new warning. The cowl and channel groups produce valid declarations for each fitted seed, with leads of 204 and 184~ms, respectively. On this event, the compression group favors window discrimination, whereas the cowl and channel groups give more consistent declaration timing across fitted seeds.

\subsection{Architecture ablations and computation}

The ablations separate scale representation from fusion (Table~\ref{tab:ablation}). InletScope has the highest mean warning $F_1$ among the three encoder variants on both changed-layout runs: 0.486 and 0.582, compared with 0.474 and 0.555 for the single-scale variant and 0.483 and 0.573 for uniform fusion. Detection favors uniform fusion at 2~mm\,s$^{-1}$ and the single-scale variant at 10~mm\,s$^{-1}$. The detection ranking therefore varies by operating condition, while learned fusion yields the highest mean warning scores on both splitter runs. The full-sensor variants share the same event-declaration outcomes: no new detection edge, and a valid warning on the 10~mm\,s$^{-1}$ event for each seed.

InletScope takes a median 1.465~ms per normalized single-sensor window in the batch-one benchmark (Table~\ref{tab:ablation}). Timing uses the detection seed-17 checkpoint, a fixed normalized 100-point input, 100 warm-up calls and 500 measured calls. Neural inference runs in evaluation mode without gradient tracking, with CUDA synchronization before and after each timed call. The measured interval includes host-to-device transfer and return of the scalar probability; RF timing includes statistical-feature extraction. Acquisition, disk access, sensor aggregation and alarm logic are excluded from this per-window measurement.

\clearpage
\begin{table}[H]
\centering
\caption{Architecture comparison on the splitter runs and batch-one computation. (a) Window $F_1$ is mean $\pm$ sample SD over three seeds. (b) Size and timings use detection seed 17, with 100 warm-up and 500 measured repetitions. Neural models use an RTX 3060; RF uses an Intel i7-14700 CPU. Time covers normalized in-memory input to host probability output, excluding acquisition, disk I/O, sensor averaging and alarm logic.}
\label{tab:ablation}
\small
\setlength{\tabcolsep}{3pt}
\begin{tabular*}{\textwidth}{@{\extracolsep{\fill}}lcccc@{}}
\toprule
\multicolumn{5}{l}{\textbf{(a) Detection and warning}} \\
\addlinespace[3pt]
 & \multicolumn{2}{c}{Detection $F_1\uparrow$} & \multicolumn{2}{c}{Warning $F_1\uparrow$} \\
\cmidrule(lr){2-3}\cmidrule(l){4-5}
Method & 2~mm\,s$^{-1}$ & 10~mm\,s$^{-1}$ & 2~mm\,s$^{-1}$ & 10~mm\,s$^{-1}$ \\
\midrule
InletScope & $0.834\pm0.003$ & $0.882\pm0.039$ & $0.486\pm0.041$ & $0.582\pm0.028$ \\
Single scale & $0.828\pm0.052$ & $0.888\pm0.032$ & $0.474\pm0.020$ & $0.555\pm0.023$ \\
Uniform fusion & $0.839\pm0.021$ & $0.861\pm0.034$ & $0.483\pm0.023$ & $0.573\pm0.037$ \\
\addlinespace[3pt]
TCN & $0.815\pm0.005$ & $0.914\pm0.003$ & $0.468\pm0.019$ & $0.486\pm0.071$ \\
RF & $0.588\pm0.049$ & $0.719\pm0.027$ & $0.213\pm0.009$ & $0.837\pm0.006$ \\
\bottomrule
\end{tabular*}
\par\medskip
\begin{tabular*}{\textwidth}{@{\extracolsep{\fill}}lrrr@{}}
\toprule
\multicolumn{4}{l}{\textbf{(b) Per-sensor computation}} \\
\addlinespace[3pt]
Method & Parameters / nodes & Median (ms)$\downarrow$ & 95th percentile (ms)$\downarrow$ \\
\midrule
InletScope & 222,085 parameters & 1.465 & 1.794 \\
Single scale & 86,467 parameters & 0.706 & 0.875 \\
Uniform fusion & 219,906 parameters & 1.434 & 1.698 \\
TCN & 66,626 parameters & 0.502 & 0.679 \\
RF (300 trees) & 321,326 nodes & 23.425 & 24.011 \\
\bottomrule
\end{tabular*}
\end{table}

\section{Discussion}

Regional comparisons link detection performance to distinct pressure responses (Fig.~\ref{fig:regions}). In the downstream-throttling run, compression-side pressure rises from 7.2 to 21.2~kPa across onset, and this region gives the highest detection $F_1$. During splitter adjustment, cowl and high-speed-channel pressures rise across onset and give higher detection scores. Compression-side pressure remains near 2.6~kPa at 2~mm\,s$^{-1}$ and decreases at 10~mm\,s$^{-1}$. The operating-path dependence observed in throat-regulation and mode-transition experiments \cite{jin2023,chen2024,zhang2026restart} provides a physical context for these regional differences. Measurement regions should therefore be assessed for the intended maneuver rather than assigned a condition-independent ranking.

Aggregation reduces sensitivity to individual sensor selection (Fig.~\ref{fig:sensors}). In both splitter runs, eight-sensor subsets have a narrower detection-score distribution than singleton subsets, with mean $F_1$ close to the full-set result. Sensor composition nevertheless affects the alarm sequence. The fixed four-sensor coverage sets give slightly higher mean detection $F_1$ than the full sets while increasing mean false-alarm episodes from 2.3 to 3.0 at 2~mm\,s$^{-1}$ and from 2.0 to 3.7 at 10~mm\,s$^{-1}$. Preserving window discrimination with fewer sensors thus does not ensure the same alarm behavior. Sensor count and regional coverage need to be assessed together over complete operating records.

The baseline and ablation comparisons (Tables~\ref{tab:main} and~\ref{tab:ablation}) show task-dependent benefits of the temporal representation. InletScope exceeds RF in mean detection $F_1$ on all three held-out runs and TCN on the downstream and slower splitter runs. TCN gives the highest mean on the faster splitter run. Learned scale fusion gives the highest mean warning $F_1$ among the encoder variants on both splitter runs, whereas single-scale or uniform-fusion variants give the highest detection means. The unchanged full-sensor event-declaration pattern across these variants shows that differences in window scores do not necessarily change the event outcome.

Continuous alarm histories connect the pressure response to the timing of the monitoring decision (Figs.~\ref{fig:timelines} and~\ref{fig:tradeoff}). On the 10~mm\,s$^{-1}$ splitter run, the compression-side group yields the highest regional warning $F_1$, while the cowl and high-speed-channel groups give consistent new-warning timing across the fitted seeds. Sensor selection therefore affects both pressure discrimination and the temporal structure of the output. Three-output confirmation suppresses brief threshold excursions and adds 20~ms to a sustained crossing. The first qualifying confirmed edge determines warning lead, while alarm duration and reset behavior describe the subsequent response. These quantities complement the 0.5-s future interval used to define the prediction target.

The seven runs contain seven annotated onsets, with three held out. Seeds and sensor subsets describe variability within these events. The 10~mm\,s$^{-1}$ warning case contains one false-alarm episode per seed over approximately 0.28~s of eligible normal exposure, making the normalized rate a short-record estimate. Assessment of sustained operation requires longer normal histories and independently acquired onset and recovery events, accompanied by run-specific calibration, acquisition-filter settings and event-timing uncertainty.

\FloatBarrier
\section{Conclusions}

InletScope was evaluated for pressure-based unstart detection and warning under changes in sensor count, measurement region and operating condition. Shared multiscale encoding and probability averaging allow each task-specific model to process different measurement sets with fixed weights and source-derived settings. The Mach-2.9 experiments support the following conclusions.

\begin{enumerate}
\item Full-sensor detection yields mean window $F_1$ scores of 0.834--0.964 across the three held-out runs. In both splitter runs, eight-sensor aggregation narrows the variation across sensor subsets while retaining a mean score close to that of the full set. It provides a tested sensor-budget option for stable window discrimination.

\item Regional detection performance depends on the operating condition. Cowl and high-speed-channel measurements yield higher detection scores during splitter adjustment, whereas compression-side measurements give the highest score during downstream throttling. The accompanying pressure histories support assessing regional coverage for each operating maneuver.

\item Complete-record evaluation identifies threshold-crossing time and alarm persistence as additional criteria for sensor configuration. Three-output confirmation suppresses brief excursions and adds 20~ms to a sustained crossing. Evaluating window discrimination together with false-alarm exposure and declaration timing connects the sensing configuration to the continuous monitoring response.
\end{enumerate}

The engineering implication is to evaluate a sensing configuration through both its pressure-classification response and its continuous alarm history. This links sensor budgeting and placement to the timing and persistence of the monitoring decision.

\section*{Acknowledgements}

This work was supported by the National Natural Science Foundation of China under Grant Nos.\ 62136004 and 62276130.

\section*{Data and code availability}

The code and synthetic example data are available at \url{https://github.com/6jm233333/InletScope}.

% References are embedded; no BibTeX step or external bibliography is required.

\end{document}